# Family tree of perovskite-related superconductors

Hans Hermann Otto

*Materialwissenschaftliche Kristallographie, TU Clausthal,*
*D-38678 Clausthal-Zellerfeld, Germany*

A simple empirical relation has been found to exist between optimum $T_c$ and the formal mean cation charge $\langle q_c \rangle$ of perovskite-related superconductors, covering both conventional superconductors and superconducting cuprates. $T_c$ is shown to increase exponentially with decreasing $\langle q_c \rangle$. It is suggested that a Ba-based cuprate with $\langle q_c \rangle \cong 2$ could reach a $T_c$ around 200 K. The strong correlation may be thought of as an indication for a common mechanism of superconductivity of the whole family of compounds. In addition, the structural assignment for the 'cubic' high-$T_c$ phase reported by Volkov ($T_c$ = 117 K) to the orthorhombic $BaCuO_{2.5}$ prototype is proposed.

## I. INTRODUCTION

Since the discovery of high-$T_c$ superconductivity in cuprates by Bednorz and Müller (1986) [1] its theoretical explanation is still unsatisfactory. However, recently there has been experimental as well as theoretical evidence that the phenomenon of superconductivity can be traced back to a common mechanism for a large group of compounds or even for all superconducting phases. Some years ago Uemura et al. (1991) [2] suggested that a large group of compounds belong to a unique class of superconductors characterized by high $T_c$ relative to $n_s$/m*, where $n_s$ is the nominal density of charge carriers and m* their effective mass. Using results of their revived RVB theory, Anderson et al. (2004) [3] proposed that super-exchange instead of phonons may be responsible for d-wave superconductivity in cuprates. Recently two independent teams (Heid et al., 2008 [4]; Giustino et al., 2008 [5]) reported that a phonon-based approach indeed failed to fully explain high-$T_c$ superconductivity. Huang [6], also questioning the BCS theory, published in 2008 a rather simple, still little recognized, unified theory of superconductivity. He argues that the real space effect of Coulomb confinement of dimerized spin-parallel electrons in stripes may be responsible for superconductivity. Experimental evidence of the presence of dynamic stripes in optimum-doped superconductors was previously given by Reznik et al. (2006) [7]. Finally, Valla et al. (2006) [8] demonstrated that the 'pseudogap' observed in the energy spectrum of unconventional superconductors may be the result of electrons bond to pairs already above the transition temperature, while superconductivity begins as result of phase coherence and collective behavior of these pairs, once $T_c$ is reached. In the present contribution it will be shown without any complicated physics that the global raising of the transition temperature $T_c$ of optimum-doped perovskite-related superconductors, covering conventional as well as unconventional superconductors, systematically depends on a decreasing mean cationic charge. This can be thought of as further indication of a common origin of superconducting properties.

## II. EMPIRICAL RELATION BETWEEN $T_c$ AND THE MEAN CATION CHARGE $\langle q_c \rangle$

An empirical relation between optimum $T_c$ and the formal mean cation charge $\langle q_c \rangle$ of perovskite-related superconductors is illustrated in Figure 1. The compounds considered cover both conventional superconductors and superconducting cuprates. Starting with the maximum cation-deficient perovskite, represented by the non-superconducting $ReO_3$ structure ($q_c$ = 6), the regression curve of $T_c$ versus $\langle q_c \rangle$ extends from the cation-deficient perovskites of the molybdenum bronze type [9,10], $BaPb_{0.75}Bi_{0.25}O_3$ [11] and $Ba_{0.6}K_{0.4}BiO_3$ [12], $(La,Sr)_2CuO_4$ [1], $YBa_2Cu_3O_{6.9}$ [13], various cuprates with cations of 6*s* electron configuration (Tl, Pb, Bi, Hg) finally to the mercury-containing cuprates with the archived highest transition temperature of $T_c$ = 164 K [14-17]. As an exception, the $LiTi_2O_4$ spinel phase ($T_c$ = 14 K ) [18] with mixed valent $Ti^{3+/4+}$ ions is included into Figure 1 showing some deviation from the curve.

On the other hand, despite its higher mean anion charge compared to oxide superconductors, the recently discovered $SmO_{1-x}F_xFeAs$ superconductor ($T_c$ = 55 K) [19] matches this curve also surprisingly well.

Since doping influences $\langle q_c \rangle$ as well as $T_c$, the reported optimum doping level to obtain the highest $T_c$ has always been used to construct Figure 1. Then the optimum of each transition temperature is connected to the mean cationic charge $\langle q_c \rangle$ by the exponential relation

$$\boldsymbol{T_c = 201 \cdot \exp\{-2.04 \cdot f(\langle q_c \rangle - 2)^{0.74}\}}, \quad (1)$$

where $T_c$ = 201 K for $\langle q_c \rangle$ = 2, and the factor $f$ formally adjusts the dimension of the argument. The equation resembles a formulation that is applicable for most sorts of interaction as well.

The deviation between observed and calculated $T_c$ is less than 9 K for a subset of 24 well-characterized data sets used from Table 2. No allocation could be made, for instance, for the very interesting interstitially doped $Ba_2Ca_{n-1+x}Cu_{n+y}O_z$ family ($T_c$ up to 126 K), because the chemical information given is insufficient [54].

In the fit the argument $\langle q_c \rangle$ -2 was applied because a cation to oxygen ratio of 1:1 is assumed to be a natural limit for both the oxygen-depleted perovskite type represented by the $Ca_{0.86}Sr_{0.14}CuO_2$ infinite-layer structure [20], and the rocksalt structure. The result suggests that a properly doped superconductor with $\langle q_c \rangle$ around 2 could reach a transition temperature of about 200 K, possibly a little more, which still awaits discovery. A maximum transition temperature of about 200 K was predicted in 1997 by Kresin et al. [21]. For too low a $\langle q_c \rangle$ the observed transition temperature already shows decreasing tendency.

The transition temperature of compounds with charge reservoir blocks, containing cations of variable valence states, is clearly enhanced in comparison to that of compounds like $La_{1-x}Sr_xCuO_4$ without such proper blocks. This fact is considered by the exponent $\alpha$ = 0.74 in eq. 1 that renders the high $T_c$ region somewhat steeper. In addition, $La^{3+}$ seems to be less efficient than the larger $Ba^{2+}$ and $Sr^{2+}$ to reach a high transition temperature.

The 'rule' for maximum $T_c$ following $T_c$(Hg) > $T_c$(Tl) >$T_c$(Bi) for the respective substance classes is connected with $q_c$ of the main cation, which develops including multi-atomic occupation from $Hg^{2+}$ and $Tl^{3+}/Cu^{1+}$ to $Bi^{3+}$. In the case of Tl-compounds the replacement of some $Ca^{2+}$ by $Tl^{3+}$ was evaluated too [30,31,32].

The right side of the plot indicates clearly the region, where superconductivity is suppressed or even impossible. The present author intended to use a simple function for the fit owing to the errors in measuring or archiving $T_c$ (onset or midpoint) and some incompleteness of chemical and physical data (cationic occupancy and valence, charge carrier concentration, $T_c$ variation with pressure, phase inhomogeneity). Of course, a better fit is possible by applying a two-stage function with a more linear tendency in the high $T_c$ region.

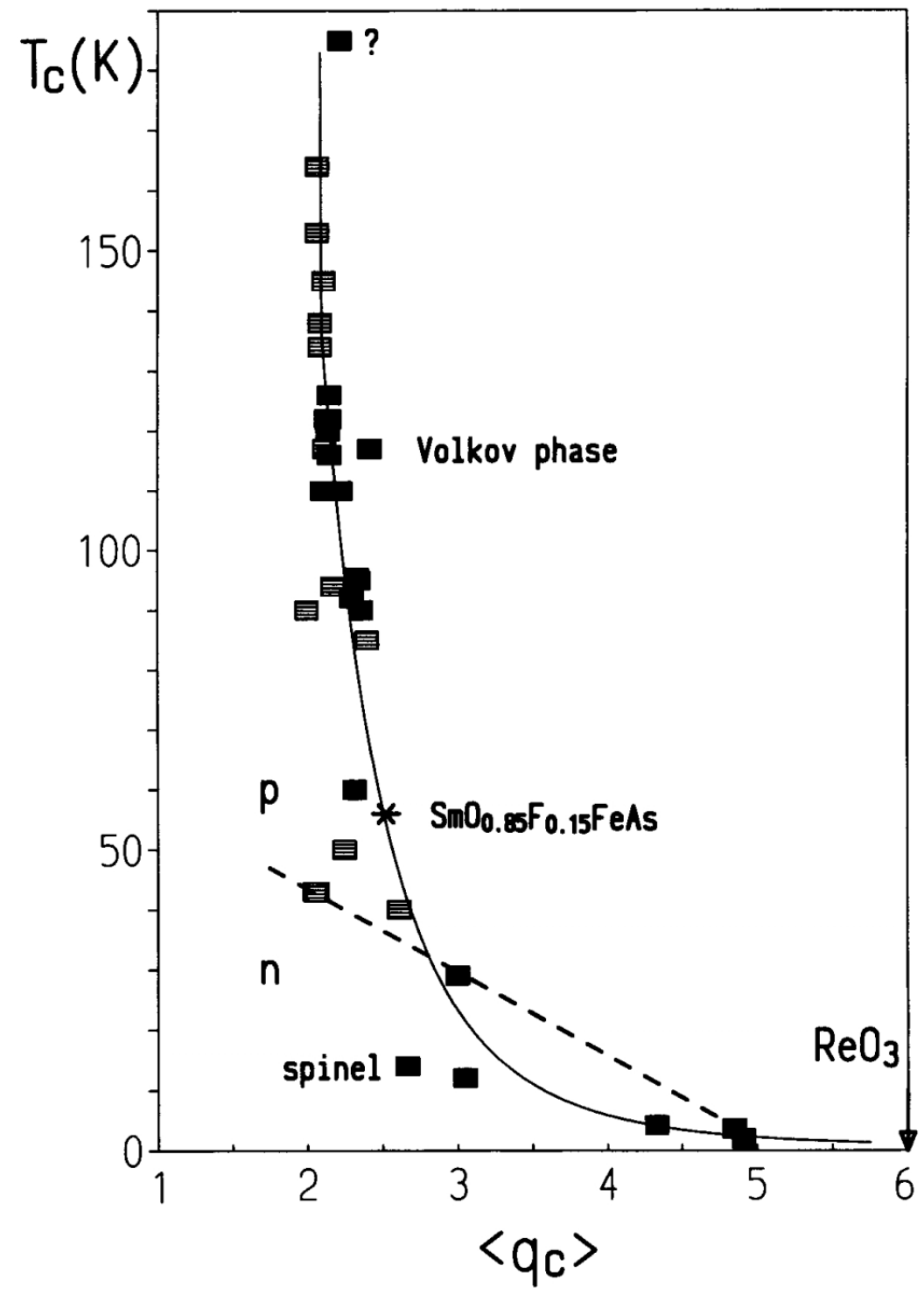

**Figure 1.** Transition temperature $T_c$ versus mean cationic charge $\langle q_c \rangle$. The assumed border between $p$ and $n$ carriers is depicted by a broken line.

The simple relation expressed by eq. 1 does not consider that superconducting properties are subject to spatial restrictions too, but because of the reason mentioned above calculation of the total energy $E_{tot} = E_M + E_b$ (electrostatic and band energy) for the entire family of compounds can not really show a more appropriate tendency.

Nonetheless, the correlation found is somewhat unexpected considering the complexity and variability of the crystal chemistry of the whole family of compounds, in which the cations and their number strongly vary and non-stoichiometry of oxygen can exist. It implies that a common underlying origin of the superconductivity of the whole class of compounds may exist, supporting the new ideas for a unified theory of superconductivity.

But what is the physics behind that relation? Decreasing $\langle q_c \rangle$ reflects an increasing cation to oxygen ratio and more evenly distributed charge of the cations. Obviously, the number of cations with low charge filling the interstices of the oxygen lattice has an exponentially increasing influence on the electronic interaction, especially on the mobility of charge carriers. An adequately low positively charged background is needed as a prerequisite for delocalization of preformed static pairs of carriers on the charge-rich stripes.

Correlations between $T_c$ and various physical parameters have been previously attempted to understand which parameters are important for an enhanced transition temperature. An inverted parabolic relation between $T_c$ and the hole concentration $n_H$, applicable to each different subclass of superconductors, was given by Wangbo and Torrardi [22], and the problem of charge distribution of holes among the various $CuO_2$ layers was considered by Di Stasio et al. [23]. Also Madelung potentials or the total energy have been calculated to understand, how dopants determine the superconducting properties [24,25]. In a recent paper Pickett reviewed variations in $T_c$ of high temperature superconductors at optimum doping levels, including the effects of underlying dispersion relation and uniaxial strain [26].

A few years ago Balasubramanian and Rao [27] plotted the transition temperature versus an electronegativity function $f_\chi = \chi_c/(\chi_c+\chi_a)$ as a more or less straight line, where the $\chi_i$ are the electronegativities for cations or anions. The function $f_\chi$ was thought to represent the pull exerted by all cations to the valence electron system. The highest $T_c$ was found for cations with the lowest formal charge and for anions with the highest one.

The plot presented by Balasubramanian and Rao [27] suffers from the scarcity of data available at that time and hence is not convincing. Even though the very simple function given in eq. 1 correlates much stronger, the result still confirms the notion that the lowest formal charge of cations leads to the highest $T_c$.

A linear relation between the critical temperature of a unique group of superconductors (cuprates, high-$T_c$ bismuthates, organic, Chevrel phase and heavy fermion systems) and the ratio $n_s/m^*$ of their nominal density $n_s$ and effective mass $m^*$ of carriers is evidenced by the universal Uemura plot [2,28]. The connection to our global $T_c$ versus $f(q_c)$ relation is indicated, if we approximate the exponential expression by the linear relation $T_c \sim \langle q_c \rangle^{-5}$. Because $n_s$ does not vary much at the optimum doping level chosen, lowering of the effective mass $m^*$, or equivalently, enhancement of the mobility of charge carriers with the inverted power of about five of mean cationic charge is again suggested as physical interpretation for the result. Time will tell whether this inverse power-law dependence can contribute to explain the microscopic mechanism of superconductivity, possibly valid for the whole perovskite tree, from low-$T_c$ bronzes to bismuthates and cuprates exhibiting highest $T_c$, from conventional superconductors to unconventional ones.

Some points have to be discussed in more detail. The Hg compounds show very large pressure derivative $dT_c/dp$ [14,15,16]. Different reasons can be responsible for this feature. Whereas $T_c$, for instance for the Hg-1223 compound measured under ambient conditions, branch off somewhat from the curve (1), the optimum $T_c$ under high pressure matches it well. Under that condition a denser phase can be re-formed in that $Hg^{2+}$ may be marginally replaced by the smaller $Cu^{1+\delta}$ in the charge reservoir layer under segregation of elementary mercury droplets, thus reducing the mean cationic charge of the superconducting phase. Indeed, one can derive from the charge derivative of eq. 1 that a marginal decrease of $\Delta q_c = 0.01$ would cause a considerable $\Delta T_c = 4.7$ K enhancement. On the other hand, Geballe et al. [29] have argued that the non-optimum off-center Hg position at ambient pressure can shift to a more favorable Hg-Hg ion separation under high pressure thus effectively creating two-site negative-U centers with then less localized electrons and smaller increase in their effective mass. However, in Tl-based superconductors, even $Tl^{3+}$ goes off-center [30,31,32], and these compounds do not show extreme pressure derivatives. Off-center positions indicate pseudo-symmetry caused by SDW behavior or twinned orthorhombic order states, which mimic tetragonal symmetry. It can be assumed that the orthorhombicity $\Delta = (b-a)/a$ of the lattice under pressure will evolve towards a value that enhances the real space collective confinement of dimerized charges in stripes. A true tetragonal lattice would indeed more hinder than favor the collective behavior of charges due to spatial frustration.

Also, the ratio of the lattice parameters of $c/b = 4.1$ under ambient conditions can be shifted under high pressure toward a rational number, easing the way to establish a Wigner supersolid. Recently such a Wigner supersolid was suggested by P. W. Anderson [33] for underdoped cuprate super-conductors, which show a '4*a* x 4*a*' superstructure. Such combined spatial features would lead to a cube with electron pairs localized at its vertices. Structural changes including proximity layer effects together with some charge redistribution or lowering of the mean cationic charge could be responsible for effective *c*-axis electrodynamics in mercury cuprates under high pressure. Since experimental evidence exists that there is no significant change in the number of charge carriers [34], pressure is thought to considerably influence the mobility of charge carriers. Lattice parameter constraints for a large group of homologous series of cuprate superconductors may be extracted from Table 1 [35,36]. Such constraints were used in Huang's unified theory of superconductivity [6]. In conclusion, both chemical and spatial reasons may work together to explain the extraordinary effect of pressure applied to Hg cuprates.

Figure 1 includes a minor 1245/1212 composite phase of the highest hitherto reported transition temperature of $T_c$ = 183 K and an assumed formula $(SnPb_{0.4}In_{0.6})Ba_4Tm_5Cu_7O_{20+}$ [85]. For the synthesis planar weight disparity was applied. Despite the difficulty to characterize such minor phase (around 1 volume-% fraction) the given formula may contain less oxygen (caused by mixed-valent $Tm^{2+/3+}$ or some $Cu^{1+}$), and the reduced mean cation charge would shift the data point towards the regression curve. Planar weight disparity, light - heavy alteration of planes within the compound, is found to enhance $T_c$ of layered perovskites. Structurally it means that cations of different size will be given the possibility to order at definite lattice sites forming a superlattice, and this may favor the self-organization of electron pairs into a Wigner supersolid. In connection with this the suggestions of Anderson [33] and Huang [6] and the seminal paper of Wigner [37] will be quoted.

## III. THE '*CUBIC*' HIGH-$T_c$ PHASE OF VOLKOV

Another point to draw attention to is the 1992 report of a 'cubic' superconducting cuprate of as yet unknown crystal structure with $T_c$ = 117 K [38] and n-type room temperature conductivity, showing, as reported, a *P* lattice with the lattice parameter of $a$ = 6.04 Å. Evidently a structure with cubic symmetry cannot be build up by infinite cuprate layers. However, the very complex formula $Tl_{0.66}Pb_{0.33}Ba_{0.71}Sr_{1.57}Ca_{2.10}Cu_{3.25}O_{10.2}F_{0.67}$ exhibits a strong similarity to the orthorhombic $BaCuO_{2.5}$ prototype [39,40] after rearranging the cations to yield $(Ba,Sr,Ca)_{4.38}(Cu,Tl,Pb)_{4.25}(O,F)_{10.87}$. Structural similarity with the 'cubic' superconductor mentioned above is further indicated by comparing the lattice parameters $a$ = 8.55 Å, $b$ = 10.56 Å, and $c$ = 7.62 Å with those calculated for three orthogonal directions of the cubic compound resulting in $a{\cdot}2^{1/2}$ = 8.54 Å, $a{\cdot}3^{1/2}$ = 10.46 Å, and $a{\cdot}(3/2)^{1/2}$ = 7.40 Å, respectively. The content of the cubic unit-cell correctly suggests that its volume would be smaller compared with $BaCuO_{2.5}$. In addition, it was reported that a Ba-rich part of the synthesis product showed splitting of reflections and larger d-values, which is consistent with our explanation of non-cubic symmetry for the superconducting main phase. Two distinct $Cu^{3+}$ sites are expected for $BaCuO_{2.5}$, one of which is localized in an octahedral ligand field with high spin configuration, and another non-octahedrally coordinated $Cu^{3+}$ atom shows low spin configuration [39]. Indeed, from the formula it can be deduced that only a part of Cu ions can exhibit octahedral coordination. For the main part of small cations high oxidation states, i.e. $Cu^{3+}$, $Tl^{3+}$ and $Pb^{4+}$ can be expected. The crystal structure may belong to a distorted perovskite type with tilted octahedra and half-octahedra. The space group, derived from the indexed powder pattern published [39], may belong to *Pnmm* or a permissible subgroup. Thus, the mystery of the unique 'cubic' high-$T_c$ phase is solved. In addition, if the proposed structural assignment can be substantiated, a new field of superconductor research could be established since it would be extremely important to know how in a crystal with such high a formal cationic charge and possibly n-type charge carriers the spin and charge stripe order may be organized. Whichever way, the data point of this phase ($\langle q_c \rangle$ = 2.42, $T_c$ = 117 K) was already used in Figure 1 to indicate its significant deviation from the other perovskite-related compounds.

## IV. THE ROLE OF LARGE CATIONS ON STRUCTURE AND PROPERTIES

The essential role played by cations with variable oxidation states, copper as well as charge reservoir cations, which are involved in the process of charge carrier creation, has been extensively analysed to unravel the many secrets of perovskite-related super-conductors. Copper oxide nets facilitate layered structures with strong interactions of electrons on the $CuO_2$ planes, but otherwise a compound with a $T_c$ as high as 117 K may not require a layered structure [38]. In the special case of $CuBa_2YCu_2O_{7-\delta}$ and the cuprates of $Hg^{2+}$ and $Tl^{3+}$, an uptake of oxygen raises the formal charge of copper ions to $Cu^{2+\delta}$ or equivalently, introduces an optimum of p-carriers (holes) in the form of $O^{1-}$ located at the centre of the charge reservoir block and on the $CuO_2$ planes next to the Ba atoms. It is of great importance that copper as $Cu^{1+}$ is able to replace some $Tl^{3+}$ or $Hg^{2+}$ in the charge reservoir block owing to the well-fitting bond distance of its dumb-bell coordination [41], thus reducing the mean cationic charge and enhancing $T_c$ in the sense of the presented relation shown in eq. 1.

On the other hand, a common feature of the chemistry of the whole family of perovskite-related superconductors is the dominant influence of large cations on structure and properties. Besides monovalent $K^{1+}$ in the superconducting bronzes and threevalent $La^{3+}$ or $Nd^{3+}$ ions in the later discovered phases [1], the divalent $Ba^{2+}$ and $Sr^{2+}$ alkaline earth ions are the most important ones, which reside on the composition plane between the different structural building blocks. The size of these strongly electropositive cations favors high oxygen coordination. From that follows a remarkable oxidation potential and the ability to accumulate some singly charged $O^{1-}$ ions in their sphere of coordination. The ease of peroxide formation ($BaO_2$, $SrO_2$) can be understood in this sense. Whereas the bond-valence sum of the peroxide matches well the valence 2+ of Ba, the bond-valence sum for the BaO rocksalt structure is underdetermined with s = 1.62+ instead of 2+, indicating that this cation is too large for a sixfold oxygen coordination. Furthermore, perovskites and related compounds often suffer from small structural distortions leading to interesting ferroic properties. Superconductivity as distinctive feature seems also to be related to a small distortion to an orthorhombic structure, which already exists at moderately high temperature. At this temperature a stripe order of charge carriers is thought to occur and a pseudogap is being opened up. We may ask whether the intrinsically existing softness of the crystal structure due to the high polarizability of oxygen and the large (soft) cations causes this charge ordering, or vice versa, the energy gain of the self-organized charge ordering is jointly responsible for the small orthorhombic distortion, where the $CuO_2$ plaquettes become inequilateral. Is it the energy gain that is responsible for self-doping of superconductors on formation, a mysterious observation that is well known to material scientists of the superconductor community?

Among all superconductors, Ba-based compounds are the most efficient with respect to highest $T_c$ (see also the remarks of Chu et al. [59]). Apart from the propensity to accumulate $O^{1-}$ the size of this cation is also responsible for suitable lattice distances obviously needed to establish an electron superlattice. Table 1 summarizes the *a* lattice parameters as well as coefficients of *c* lattice parameter relations for individual homologous series of perovskite related superconductors [35,36]. The $d_2$ = 2.80 Å spacing of the rocksalt block around $Ba^{2+}$ appears to be essential in combination with the charge reservoir separation of $d_1$ = 1.97 Å. The smaller separation around $Sr^{2+}$ ($d_2$ = 2.39 Å) works well only in combination with $Bi^{3+}$ (or $Pb^{2+}$) as charge reservoir cation, because the lone electron pair of $Bi^{3+}$ needs more space. Its dipole moment can be compensated only in a double layer arrangement, giving $d_1$ = 2.48 Å for the charge reservoir block. So $d_1$ and $d_2$ add to 4.87 Å for Bi-Sr compounds, only slightly larger than the 4.77 Å combined distance for Ba compounds. Furthermore, this distance is related to the $d_3$ separation of the infinite layers as $2(d_1 + d_2) \approx 3{\cdot}d_3$.

It might be that the exceptional role of three-layer superconductors has its origin partly in this distance relation. Between the *a* lattice parameter and the combined $d_1 + d_2$ distance the relation $4(d_1 + d_2) \approx 5{\cdot}a$ exists.

It would not be purely coincidental that charge carriers like this geometrical pattern when they begin to order. Peculiarities of superconductor phase diagrams such as magic phase locking behavior [42], sometimes referred to as devil's staircase-like [43,44], can be attributed to this spatial feature.

The importance of the distance structurally equivalent layers are separated from each other points to effects of interlayer coupling [45]. It would be tempting to assume that the next neighbouring layer knows about the collective electrodynamics in the first one. Then the picture emerges of electron pairs (not single electrons) resonating on ‘strings’ along [001]. If the large cation dependent separation and the carrier density are favourable, this allows for a supercurrent to flow through the lattice.

These remarks show that apart from copper in its different valence states and charge reservoir cations also large alkaline earth cations take an ‘active’ part in superconductivity. Therefore, it is well founded to use all cations in our consideration, but a weighting of the less active ionic part of the inner infinite layer block presently included may be useful.

## V. CONCLUSIONS

An empirical exponential relation between $T_c$ and the formal mean cationic charge $\langle q_c \rangle$ of the global class of perovskite-related superconductors suggests a common origin of the mechanism leading to superconductivity. The slope of the graph suggests for $\langle q_c \rangle \cong 2$ a possible $T_c$ of about 200 K or slightly higher for a compound hopefully to be discovered in the near future. Owing to its properties the soft $Ba^{2+}$ ion, which obviously likes $O^{1-}$ in its surrounding, is a most suitable constituent to reach this intention. The global relation described in this contribution may serve as a guide for experimentalists to further optimize superconducting properties. However, to keep down physical expectations it may be deduced that a room temperature superconductor will hardly be found among compounds of the known perovskite family, unless the Volkov phase establishes a new promising branch of the family tree.

**Acknowledgement** I am much indebted to my colleague Prof. Robert B. Heimann for his suggestions and the improvement of the manuscript.

---

[1] J. G. Bednorz, and K. A. Müller, Z. Phys. B **64**, 189 (1986).

[2] Y. J. Uemura, L. P. Le, G. M. Luke, B. J. Sternlieb, W.D. Wu, J. H. Brewer, T. M. Riseman, C. L. Seaman, M. B. Maple, M. Ishikawa, D. G. Hinks, J. D. Jorgensen, G. Saito, and H. Yamochi, Phys. Rev. Lett. **66,** 2665 (1991).

[3] P. W. Anderson, P. A. Lee, M. Randeria, T. M. Rice, N. Trivedi, and F. C. Zhang, arXiv:cond-mat/0311467v2 (2004)
[4] R. Heid, K.-P. Bohnen, R. Zeyher, and D. Manske, Phys. Rev. Lett. **100**, 137001 (2008).
[5] F. Guistino, M. L. Cohen, and S. G. Louie, Nature **452**, 975 (2008).
[6] X. Huang, arXiv:physics.gen-ph 0804.1615v1 (2008).
[7] D. Reznik, L. Pintschovius, M. Ito, S. Iikubo, M. Sato, H. Goka, M. Fujita, K. Yamada, G. D. Gu, and J. W. Tranquada, Nature **440**, 1170 (2006).
[8] T. Valla, A.V. Fedorov, J. Lee, J. C. Davis, G. D. Gu, Sciente **314**, 1914 (2006).
[9] A. R. Sweedler, Ch. J. Raub, and B. T. Matthias, Phys. Lett. **15**, 108 (1965).
[10] A. W. Sleight, T. A. Bitter, and P. E. Bierstedt, Sol. State Comm., **7**, 299 (1969).
[11] A. W. Sleight, I. L. Gillon, and P. E. Bierstett, Sol. State Comm. **17**, 27 (1975).
[12] L. F. Matheiss, E. M. Gyorgy, and D. W. Johnston Jr, Phys. Rev. B**37**, 3745 (1988).
[13] M. K. Wu, J. R. Ashburn, C. J. Torng, P. H. Hor, R. L. Meng, L. Gao, Z. J. Huang, Y. Z. Wang, and C. W. Chu, Phys. Rev. Lett. 5**8**, 908 (1987).
[14] C. W. Chu, L. Gao, F. Chen, Z. J. Huang, R. L. Meng, and Y. Y. Xue, Nature **365**, 323 (1993).
[15] M. Nuñez-Regueiro, J.-L. Tholence, E. V. Antipov, J.-J. Capponi, and M. Marezio, Science **262**, 97 (1993).
[16] L. Gao, Y. Y. Xue, F. Chen, Q. Xiong, R. L. Meng, D. Raminez, and C. W. Chu, Phys. Rev. B**50**, 4260 (1994).
[17] X. S. Wu, H. M. Shao, S. S. Jiang, C. Gou, D. F. Chen, D. W. Wang, and Z. H. Wu, Physica C **261**, 189 (1996).
[18] D. J. Johnston, J. Low Temp. Phys. **25**, 145 (1976).
[19] X. H. Chen, Z. Li, D. Wu, G. Li, W. Z. Hu, H. Chen, and D. F. Fang, cond-mat./0803.3603.
[20] T. Siegrist, S. M. Zahurak, D. W. Murphy, and R. S. Roth, Nature **334**, 231 (1988).
[21] V. Z. Kresin, S. A. Wolf, Yu. N. Ovchinnikov, Phys. Reports **288**, 347 (1997).
[22] M.-H. Wangbo and C. C. Torrardi, Science **249,** 1143 (1990)
[23] M. Di Stasio, K. A. Müller, and L. Pietronero, Phys. Rev. Lett. **64**, 2827 (1990).
[24] L. F. Feiner and D. M. de Leeuw, Sol. State Comm. **70**, 1165 (1989).
[25] R. V. Lutchiv, Y. V. Boyko, Cond. Mat. Phys. **2**, 481 (1999).
[26] W. E. Pickett, Iran. J. Phys. Res. **6**, 29 (2006).
[27] S. Balasubramanian, and K. J. Rao, Sol. State Comm. 71, 979 (1989).
[28] C. Paracchini, L. Romano, and S. Bellini, Physica C **192**, 443 (1992).
[29] T. H. Geballe, B. Y. Moyzhes, and P. H. Dickinson, arXiv:cond-mat/9904435 (1999).
[30] H. H. Otto, T. Zetterer, and K.-F. Renk, Naturwiss. **75**, 509 (1988).
[31] H. H. Otto, T. Zetterer, and K.-F. Renk, Z. Phys. B **75**, 433 (1989).
[32] T. Hertlein, H. Burzlaff, H. H. Otto, T. Zetterer, and K.-F. Renk, Naturwiss. **76**, 170 (1989).
[33] P. W. Anderson, arXiv:cond-mat/0406038 (2004).
[34] M. T. D. Orlando, E. V. L. de Mello, C. A. C. Passos, M. R. C. Caputo, L. G. Martinez, B. Zeini, E. S. Yugue, W. Vanoni, and E. Baggio-Saitovitch, Physica C **364-365**, 350 (2001).
[35] H. H. Otto, "New superconducting materials". Lectures at the TU Clausthal, 1992 to 2002.
[36] R. Baltrusch, Dissertation TU Clausthal 1997.
[37] E. P. Wigner, Phys. Rev. **46**, 1002 (1934).
[38] V. E. Volkov, A. D. Vasiliev, Yu. G. Kovalev, S. G. Ovchinnikov, N. P. Fokina, V. K. Chernov, and K. S. Aleksandrov, Pisma Zh. Eksp. Teor. Fiz. **55**, 591 (1992).
[39] M. Arjomand and D. J. Machin, J. Chem. Soc. Dalton Trans. 1061 (1975).
[40] W. Mingmei, S. Qiang, H. Gang, R. Yufang, and W. Hongyiang, J. Solid State Chem. **110**, 389 (1994).
[41] H. H. Otto, R. Baltrusch, and H.-J. Brand, Physica C **215**, 205 (1993).
[42] A. R. Moodenbaugh, Y. Xu, M. Suenaga, T. J. Folkerts, and R. N. Shelton, Phys. Rev. B **38**, 4596 (1988).
[43] J. von Boehm, P. Bak, Phys. Rev. Lett. **42**, 122 (1979)
[44] C. W. Chu, Y. Y. Xue, Y. Y. Sun, R. L. Meng, Y. K. Tan, L. Gao, Z. J. Huang, J. Bechthold, P. H. Hor, and Y. C. Jean, Proc. Taiwan Int. Symp. on Supercond. (1989).
[45] P. W. Anderson, Science **279**, 1196 (1998).
[46] S. N. Putilin, E. V. Antipov, O. Chmaissem, and M. Marezio, Nature **362**, 226 (1993).
[47] A. Schilling, M. Cantoni, J. D. Guo, and H. R. Ott, Nature **363**, 56 (1993).
[48] R. J. Cava, B. Batlogg, J. J. Krajewski, R. Farrow, L. P. Rupp Jr, A. E. White, K. Short, W. F. Peck, and T. Kometani, Nature **332**, 814 (1988).
[49] H. Takagi, S. Ushida, and Y. Tokura, Phys. Rev. Lett. **62**, 1197 (1987).
[50] J. Akimitsu, S. Suzuki, M. Watanabe, and H. Sawa, Jpn. J. Appl. Phys. **27**, L1859 (1988).
[51] M. G. Smith, A. Manthiram, J. Zhou, J. B. Goodenough, and J. T. Mackert, Nature **351**, 549 (1991).
[52] G. Er, Y. Miyamoto, F. Kanamura, and Kikkawa, Physica C **181**, 206 (1991).
[53] M. Azuma, Z. Hiroi, M. Takano, Y. Bando, and Y. Takeda, Nature **356**, 775 (1992).
[54] M. Takano, M. Azuma, Z. Hiroi, and Y. Bando, Physica C **176**, 441 (1991).
[55] Y. Shimakawa, J. D. Jorgensen, J. F. Mitchell, B. A. Hunter, H. Shaked, D. G. Hinks, R. L. Hittermann, Z. Hiroi, and M. Takano, Physica C **28**, 73 (1994).
[56] J. P. Hodges, P. R. Slater, P. P. Edwards, C. Greaves, M. Slaski, G. Van Tendeloo, and S. Amelincks, Physica C **260**, 240 (1996).
[57] R. J. Cava, B. Batlogg, R. van Dover, J. J. Krajewski, J. V. Waszczak, R. M. Fleming, W. F. Peck Jr, L. W. Rupp Jr, P. Marsh, C. W. P. James, and L. F. Schneemeyer, Nature **345**, 602 (1990).
[58] Z. Hiroi, M. Tanako, M. Azuma, and Y. Takeda, Nature **364**, 315 (1993).

[59] C. W. Chu, Y. Y. Xue, Z. L. Du, Y. Y. Sun, L. Gao, N. L. Wu, Y. Cao, I. Rusakova, K. Ross, Science **277**, 1081 (1997).
[60] S.S.P. Parkin, V. Y. Lee, A. I. Nazzal, R. Savoy, T. C. Huang, G. Gorman, and R. Beyers, Phys. Rev. B **38**, 6531 (1988).
[61] S. Li, M. Greenblatt, and A. J. Jacobson, Mater. Res. Bull. **26**, 229 (1991).
[62] P. Haldar, S. Sridhar, S. Roig-Janicki, W. Kennedy, D. H. Wu, C. Zahopoulos, and B. C. Giessen, J. Supercond. **1**, 211 (1988).
[63] R. L. Meng, Y. Y. Sun, J. Kulik, Z. J. Huang, F. Chen, Y. Y. Xue, and C. W. Chu, Physica C **214**, 307 (1993).
[64] T. Maeda, N. Sakayuma, S. Koriyama, A. Ichinose, H. Yamauchi, and S. Tanaka, Physica C **169**, 133 (1990).
[65] Z. Y. Chen, Y. Q. Tang, Y. F. Li, D. O. Pederson, and Z. Z. Sheng, Mat. Res. Bull. **27**, 1049 (1992).
[66] Y. Miyazaki, H. Yamane, N. Kobayashi, T. Hirai, H. Nakata, K. Tomimoto, and S. Akimitsu, Physica C **202**, 162 (1992).
[67] C. Martin, C. Michel, A. Maignan, M. Hervieu, and B. Raveau, C. R. Acad. Sci. Ser.2, **307**, 27 (1988).
[68] C. Martin, M. Huve, M. Hervieu, A. Maignan, C. Michel, and B. Raveau, Physica C **201**, 362 (1992).
[69] M. A. Subramanian, C. C. Torardi, J. Gopalakrishnan, P. L. Gai, J. C. Calabrese, T. R. Askew, R. B. Flippen, and A. W. Seight, Science **242**, 249 (1988).
[70] L. Gao, Z. L. Du, Y. CaO, I. Rusakova, Y.Y. Sun, Y. Y. Xue, and C. W. Chu, Mod. Phys. Lett. B**9**, 1397 (1995).
[71] I. Bryntse and A. Kureiva, Mat. Res. Bull. **30**, 1207 (1995).
[72] H. Ihara, R. Sugise, M. Hirabayashi, N. Terada, M. Jo, K. Hayashi, A. Negishi, M. Tokumoto, Y. Kimura, and T. Shimomura, Nature **334**, 510 (1988).
[73] E. V. Antipov, S. M. Loureiro, C. Chaillout, J. J. Capponi, P. Bordet, J. L. Tholence, S. N. Putilin, and M. Marezio, Physica C **215**, 1 (1993).
[74] J. Akimoto, Y. Oosawa, K. Tokiwa, M. Hirabayashi, and H. Ihara, Physica C **242**, 360 (1995).
[75] H. Ihara, K. Tokiwa, H. Ozawa, M. Hirabayashi, H. Matuhata, A. Negishi, and Y. S. Song, Jpn. J. Appl. Phys. **33**, L300 (1994).
[76] C. C. Torardi, M. A. Subramanian, J. C. Calabrese, J. Gopalakrishnan, E. M. MaCarron, K. J. Morrisey, T. R. Askew, R. B. Flippen, U. Chawdhry, and A.W. Sleight, Phys. Rev. B **38**, 225 (1988).
[77] J. B.Parise, N. Herron, M. K. Crawfford, P. L. Gai, Physica C **159**, 225 (1989).
[78] C. Michel, M. Hervieu, M. M. Borel, A. Grandin, F. Deslandes, J. Provost, and B. Raveau, Z. Phys. B **68**, 421 (1987).
[79] P. Marsh, R. M. Fleming, M. L. Mandich, A. M. DeSantolo, J. Kwo, M. Hong, and L. J. Martinez-Miranda, Nature **334**, 141 (1988).
[80] J. Gopalakrishnan, C. Shivakumara, and V. Manivannan, Mat. Res. Bull. **29**, 369 (1994).
[81] H. Maeda, Y. Tanaka, M. Fukutomi, and T. Asano, Jpn. J. Appl. Phys. **27**, 1209 (1988).
[82] C. C. Torardi, M. A. Subramanian, J. C. Calabrese, J. Gopalakrishnan, E. M. McCarron, K. J. Morrisey, T. R. Askew, R. B. Flippen, U. Chowdhry, and A. W. Sleight, Science **240**, 631 (1988).
[83] J. M. Tarascon, Y. LePage, P. Barboux, P. G. Bagley, L. H. Greene, W. R. McKinnon, G. W. Hull, M. Giroud, and D. M. Hwang, Phys. Rev. B **37**, 9382 (1988).
[84] M. R. Presland, J. L. Tallon, P. W. Gilberd, and R. S. Liu, Physica C **191**, 307 (1992).
[85] E. J. Eck, Superconductor News, Superconductors.ORG 2008.

Table 1. Lattice parameter relations of important homologous series of superconductors deduced from structural building blocks [35,36]. The *c* lattice parameter is given as $\mathbf{c} = \mathbf{n_L}[\,(\mathbf{n_1}+1)d_1 + \mathbf{n_2}d_2 + \mathbf{n_3}d_3]$, where the $\mathbf{n_i}$ are the cation numbers of the four-membered formula symbol and $\mathbf{n_L} = 1$, if $\sum \mathbf{n_i}$ = even, i = 1 to 4 (*P*-lattice), $\mathbf{n_L} = 2$, if $\sum \mathbf{n_i}$ = odd (*I*-lattice, *A*-lattice). $(4a^2d_2)^{1/3}$ is comparable with the lattice parameter of BaO or SrO (rocksalt-type).

| **Homologous series** | | | ***a*** (Å) | $\mathbf{d_1}$(Å) | $\mathbf{d_2}$(Å) | $\mathbf{d_3}$(Å) | $\mathbf{(4a^2d_2)^{1/3}}$(Å) |
|---|---|---|---|---|---|---|---|
| $\mathbf{n_1}$ | $\mathbf{n_2}$ | $\mathbf{n_3}$ | | charge reservoir | rocksalt | infinite layer | |
| Cu | Ba | Y | 3.85 | 1.90 | 2.53 | 2.86 | 5.31 |
| Tl | Ba | Ca | 3.85 | 1.98 | 2.80 | 3.14 | 5.50 |
| Hg | Ba | Ca | 3.86 | 1.96 | 2.80 | 3.17 | 5.51 |
| Hg | Ba | Y(Ln) | 3.88 | 1.96 | 2.80 | 3.00 | 5.52 |
| - | Ba | Ca | 3.85 | 1.96 | 2.80 | 3.16 | 5.50 |
| - | (La,Sr) | Ca | 3.80 | 1.81 | 2.41 | 3.16 | 5.18 |
| (Tl,Pb) | Sr | Ca | 3.80 | 2.08 | 2.39 | 3.15 | 5.17 |
| Bi | Sr | Ca | 3.80 | 2.48 | 2.39 | 3.16 | 5.17 |
| Cu/Pb | Sr | Y | 3.83 | 1.93 | 2.44 | 3.13 | 5.23 |
| (Pb,Tl) | Sr | Tl | 3.84 | 1.98 | 2.35 | 2.98 | 5.18 |

Table 2. Data of perovskite-related superconductors (* see page 4).
Lattice parameters *a* represent in some cases reduced values $a' = a/\sqrt{2}$.

| **Formula** | **Symbol** | ***a***(Å) | ***b***(Å) | ***c***(Å) | $T_c$ (K) | Reference | |
|---|---|---|---|---|---|---|---|
| $Rb_{0.28}WO_3$ (hexagonal) | 0 1 0 1 | 7.4 | | 7.6 | 2.0 | Sweedler et al., 1965 | [9] |
| $K_{0.3}ReO_3$ (hexagonal) | 0 1 0 1 | 7.335 | | 7.48 | 3.6 | Sleight et al., 1969 | [10] |
| $K_{0.5}MoO_3$ (tetragonal) | 0 1 0 1 | 12.36 | | 3.86 | 4.2 | Sleight et al., 1969 | [10] |
| $BaPb_{0.75}Bi_{0.25}O_3$ | 0 1 0 1 | 4.303 | | | 19 | Sleight et al., 1975 | [11] |
| $Ba_{0.60}K_{0.40}BiO_{3-\delta}$ | 0 1 0 1 | 4.293 | | | 29 | Mattheis [12]; Cava | [48] |
| $(Nd,Ce)_2CuO_4$ | 0 0 2 1 | 3.96 | | 12.08 | 29 | Takagi et al., 1987 | [49] |
| $(Nd,Sr)(Nd,Ce)CuO_4$ | 0 1 1 1 | 3.856 | | 12.48 | 30 | Akimitsu et al., 1988 | [50] |
| $Sr_{0.86}Nd_{0.14}CuO_2$ | 0 0 1 1 | 3.942 | | 3.38 | 40 | Smith et al., 1991 | [51] |
| $Sr_{0.9}La_{0.1}CuO_2$ | 0 0 1 1 | | | | 43 | Er et al., 1991 | [52] |
| $(Sr_{0.7}Ca_{0.3})_{0.9}CuO_2$ | 0 0 1 1 | 3.902 | | 3.35 | 110 | Azuma et al., 1992 | [53] |
| $Ba_{0.2}Sr_{0.8}CuO_2$ | 0 0 1 1 | | | | 90 | Takano et al., 1991 | [54] |
| $La_{1.84}Sr_{0.16}CuO_4$ | 0 2 0 1 | 3.79 | 3.80 | 13.25 | 40 | Bednorz, Müller, 1986 | [1] |
| $Sr_2CuO_{3+\delta}$ | 0 2 0 1 | 3.764 | | 12.55 | 70 | Shimakawa et al., 1994 | [55] |
| $Ba_{1.2}Sr_{0.8}CuO_{3+\delta}$ | 0 2 0 1 | 3.899 | | 12.82 | 50 | Hodges et al., 1996 | [56] |
| $(La,Sr)_2CaCu_2O_6$ | 0 2 1 2 | 3.82 | | 19.60 | 60 | Cava et al., 1990 | [57] |
| $Sr_3Cu_2O_{5+\delta}$ | 0 2 1 2 | 3.902 | | 21.09 | 100 | Hiroi et al., 1993 | [58] |
| $Ba_2Ca_{2+x}Cu_{3+y}O_z$ | 0 2 2 3 | 3.850 | | 28.2 | 126 | Chu et al. (1997) | [59] |
| $Ba_2Ca_{3+x}Cu_{4+y}O_z$ | 0 2 3 4 | 3.850 | | 34.8 | 117 | Chu et al. (1997) | [59] |
| $Tl\ Ba_2CuO_5$ | 1 2 0 1 | 3.830 | | 9.55 | <10 | Parkin et al., 1988 | [60] |
| $(Tl,Bi)Sr_2CuO_5$ | 1 2 0 1 | 3.745 | | 9.00 | 50 | Li et al., 1991 | [61] |
| $HgBa_2CuO_{4+\delta}$ | 1 2 0 1 | 3.880 | | 9.51 | 94 | Putilin et al., 1993 | [46] |
| $CuBa_2YCu_2O_{6.9}$ | 1 2 1 2 | 3.82 | 3.88 | 11.68 | 92 | Wu et al., 1987 | [13] |
| $TlBa_2CaCu_2O_7$ | 1 2 1 2 | 3.833 | | 12.68 | 90 | Parkin et al., 1988 | [60] |
| $(Tl,Pb)Sr_2CaCu_2O_7$ | 1 2 1 2 | 3.794 | | 12.08 | 85 | Subramanian et al., 1988 | [69] |
| $(Tl,Bi)Sr_2CaCu_2O_7$ | 1 2 1 2 | 3.80 | | 12.07 | 90 | Haldar et al., 1988 | [62] |
| $HgBa_2CaCu_2O_{6+\delta}$ | 1 2 1 2 | 3.862 | | 12.71 | 117 | Meng et al., 1993 | [63] |
| $(Pb,Cu)(Sr,Eu)_2(Eu,Ce)_2Cu_2O_{9-\delta}$ | 1 2 2 2 | 3.80 | | 29.60 | 25 | Maeda et al., 1990 | [64] |
| $(Tl,Pb)Sr_2(Nd,Ce)_2Cu_2O_9$ | 1 2 2 2 | 3.878 | | 30.42 | 40 | Chen et al., 1992 | [65] |
| $(C,Cu)Sr_2(Y,Ce)_2Cu_2O_x$ | 1 2 2 2 | 3.827 | | 27.71 | 18 | Miyazaki et al., 1992 | [66] |
| $Tl\ Ba_2Ca_2Cu_3O_9$ | 1 2 2 3 | 3.848 | | 15.89 | 120 | Martin et al., 1988 | [67] |
| $Tl\ Sr_2Ca_2Cu_3O_9$ | 1 2 2 3 | 3.832 | | 15.59 | 116 | Martin et al., 1992 | [68] |
| $(Tl,Pb)Sr_2Ca_2Cu_3O_9$ | 1 2 2 3 | 3.808 | | 15.23 | 122 | Subramanian et al., 1988 | [69] |
| $(Ca,Cu,Ag)Ba_2Ca_2Cu_3O_{8+\delta}$ | 1 2 2 3 | 3.8 | | | 124 | Gao et al., 1995 | [70] |
| $HgBa_2Ca_2Cu_3O_{8+\delta}$ | 1 2 2 3 | 3.86 | | 15.90 | 134 | Schilling et al., 1993 | [47] |
| $HgBa_2Ca_2Cu_3O_{8+\delta}$ | 1 2 2 3 | 3.856 | 3.838 | 15.851 | 134 | Bryntse, Kureiva, 1995 | [71] |
| $HgBa_2Ca_2Cu_3O_{8+\delta}$ | 1 2 2 3 | | | | 153 | Chu et al., 1993 | [14] |
| $HgBa_2Ca_2Cu_3O_{8+\delta}$ | 1 2 2 3 | | | | 150 | Nuñez-Regueiro et al. | [15] |
| $HgBa_2Ca_2Cu_3O_{8+\delta}$ | 1 2 2 3 | | | | 164 | Gao et al., 1994 | [16] |
| $Hg_{0.7}Pb_{0.3}Ba_2Ca_2Cu_3O_{8.45}$ | 1 2 2 3 | 3.846 | | 15.825 | 143 | Wu et al., 1996 | [17] |
| $TlBa_2Ca_3Cu_4O_{11}$ | 1 2 3 4 | 3.838 | | 18.98 | 122 | Ihara et al., 1988 | [72] |
| $HgBa_2Ca_3Cu_4O_{10+\delta}$ | 1 2 3 4 | 3.854 | | 19.01 | 126 | Antipov et al., 1993 | [73] |
| $Cu_{0.6}Ba_2Ca_3Cu_4O_{10.8}$ | 1 2 3 4 | 3.853 | | 17.97 | 116 | Akimoto et al., 1995 | [74] |
| $Ag_{1-x}Cu_xBa2Ca3Cu_4O_{11-\delta}$ | 1 2 3 4 | 3.864 | | 18.11 | 117 | Ihara et al., 1994 | [75] |
| $(Tl,Cu)_2Ba_2CuO_6$ | 2 2 0 1 | 3.866 | | 23.23 | 90 | Torardi et al., 1988 | [76] |
| $(Tl,Cd)_2Ba_2CuO_6$ | 2 2 0 1 | 3.851 | | 23.32 | 92 | Parise et al., 1989 | [77] |
| $Bi_2(Sr,Ca)_2CuO_{6+\delta}$ | 2 2 0 1 | 3.823 | | 30.77 | 10 | Michel et al., 1987 | [78] |
| $Cu_2Ba_2YCu_2O_8$ | 2 2 1 2 | 3.857 | | 27.24 | 80 | Marsh et al., 1988 | [79] |
| $(Tl,Cu)_2Ba_2(Ca,Tl)Cu_2O_8$ | 2 2 1 2 | 3.856 | | 29.30 | 110 | Gopalakrishnan, 1994 | [80] |
| $Bi_2Sr_2CaCu_2O_{8+\delta}$ | 2 2 1 2 | 3.823 | | 30.77 | 95 | Maeda et al., 1988 | [81] |
| $(Tl,Cu)_2Ba_2(Ca,Tl)_2Cu_3O_{10}$ | 2 2 2 3 | 3.850 | | 35.64 | 126 | Torardi et al., 1988 | [82] |
| $Bi_2Sr_2Ca_2Cu_2O_{10+\delta}$ | 2 2 2 3 | 3.818 | | 37.10 | 110 | Tarascon et al., 1988 | [83] |
| $(Tl,Cu)_2Ba_2(Ca,Tl)_3Cu_4O_{12}$ | 2 2 3 4 | 3.852 | | 41.99 | 116 | Presland et al., 1992 | [84] |
| $(Ba,Sr,Ca)_{4.38}(Cu,Tl,Pb)_{4.25}(O,F)_{10.87}$ | | 6.04 * | | | 117 | Volkov et al., 1992; | [38] |